\documentclass[traditabstract]{aa}
\bibpunct{(}{)}{;}{a}{}{,} 
\pdfoutput=1
\usepackage{graphicx}
\usepackage[varg]{txfonts}
\usepackage{color}
\usepackage{lscape}
\usepackage{pdfpages}

\newcommand{\srm}{\scriptscriptstyle\rm}

\begin{document}

\title{Optical linear polarization measurements of quasars obtained with the Very Large Telescope at Paranal Observatory.\thanks{Based on observations made with the ESO VLT at the Paranal Observatory under programme ID 081.A-0023, 082.B-0029, 095.A-0600}}
\author{D. Hutsem\'ekers\inst{1,}\thanks{Senior Research Associate F.R.S.-FNRS},
        B. Borguet\inst{1},
        D. Sluse\inst{1},
        V. Pelgrims\inst{2}
        }
\institute{
    Institut d'Astrophysique et de G\'eophysique,
    Universit\'e de Li\`ege, All\'ee du 6 Ao\^ut 19c, B5c,
    4000 Li\`ege, Belgium
    \and
    Laboratoire de Physique Subatomique et de Cosmologie, 53 Avenue des Martyrs, 38026 Grenoble, France
    }
\date{Received ; accepted: }
\titlerunning{Quasar polarization measurements} 
\authorrunning{D. Hutsem\'ekers et al.}
\abstract{We report 87 previously unpublished optical linear polarization measurements of 86 quasars obtained in May and October 2008, and from April to July 2015 with the Focal Reducer and low-dispersion Spectrographs FORS1 and FORS2 attached to the Very Large Telescope at the Paranal Observatory. Of the 86 quasars, 37 have $p \geq 0.6\%$, 9 have $p \geq 2\%$, and 1 has $p \geq 10\%$.
}
\maketitle
%
%
%
\section{Introduction}
\label{sec:intro}

The linear polarization of optical light is an important element in the study of quasars and other active galatic nuclei (AGN).  Polarization is a key feature in AGN unification models since it provides a periscopic view of the AGN core \citep{1985Antonucci,1993Antonucci,2005Zakamska}. Several types of quasars, such as blazars, broad absorption line (BAL) quasars, and red quasars \citep{1981Moore, 1984Stockman, 1998bHutsemekers, 2018Alexandroff}, are characterized by a polarization higher than normal, which reveals specific physical processes (jets, outflows, etc). When attributed to scattering, polarization is related to the object symmetry axis, even if the object is spatially unresolved. Correlated to the object morphology, polarization allows us to study alignments of distant quasars with the large-scale structures in which they are embedded \citep{2005Hutsemekers,2014Hutsemekers}. Polarization is also sensitive to tiny physical effects that can be revealed by the huge travel distance of photons that are emitted by quasars, and it is therefore a useful tool for testing cosmic birefringence that is caused by departures from the Einstein equivalence principle or by hypothetical axion-like particles, among others \citep{2010DiSerego,2015DiSerego,2012Payez}.

We here report new optical linear polarization measurements of quasars obtained with the FOcal Reducer and low-dispersion Spectrographs FORS1 and FORS2 attached to the Very Large Telescope (VLT) installed at the Paranal European Southern Observatory (ESO). The observations were designed for different scientific goals, but the quality of the data is homogeneous.

In Sect.~\ref{sec:obs} we outline the observing procedure. Data reduction and measurements are summarized in Sect.~\ref{sec:reduc}. The online table with the final measurements is described in Sect.~\ref{sec:data}.

\section{Observations}
\label{sec:obs}

The polarimetric observations were carried out at the ESO VLT using FORS1 in May 2008 (visitor mode) and October 2008 (service mode), and with FORS2 from April to July 2015 (service mode). Both FORS1 and FORS2 were mounted at the Cassegrain focus of the Unit Telescopes (UTs). Linear polarimetry was performed by inserting a Wollaston prism into the parallel beam. This prism splits the incoming light rays into two orthogonally polarized beams. Each object in the field therefore has two orthogonally polarized images on the CCD detector, separated by 22$\arcsec$. To avoid image overlapping, the Multi-Object Spectroscopy (MOS) slits were used to create a mask of alternating transparent and opaque parallel strips whose widths correspond to the splitting.  The final CCD image consists of alternate orthogonally polarized strips of the sky, two of them containing the polarized images of the object itself.  Because the two orthogonally polarized images of the object are recorded simultaneously, the polarization measurements do not depend on variable atmospheric transparency or seeing. In order to derive the normalized Stokes parameters $q$ and $u$, four frames were obtained with the half-wave plate (HWP) at four different position angles (0$\degr$, 22.5$\degr$, 45$\degr$, and 67.5$\degr$). Two different orientations of the HWP are sufficient to measure the linear polarization, but the two additional orientations allowed us to remove most of the instrumental polarization. The targets were positioned at the center of the field to avoid the significant off-axis instrumental polarization generated by the FORS optics \citep{2006Patat}. In May 2008, polarized and unpolarized standard stars were observed to check the zero-point of the polarization position angle, to estimate the residual instrumental polarization, and to verify the whole observing and reduction process. Standard stars were also occasionally observed in service mode (Table~{\ref{tab:std}}).

Most targets are quasars with redshifts between one and three and V magnitudes between 17.5 and 19.5. All but two (3C138 and SDSSJ204727.54-052118.8) are at high galactic latitudes $|b_{\rm gal}| > 30 \degr$. They were mostly selected among broad absorption line, radio-loud, or red quasars that are more likely to be significantly polarized, or among quasars belonging to large quasar groups \citep{2014Einasto,2014Hutsemekers}. All observations were obtained through the V-high$+114$ filter with typical exposure times per frame ranging between 30 seconds and 10 minutes. The FORS1 CCD was a 4k$\times$4k E2V mosaic, used with binning 1$\times$1 in May 2008 and  2$\times$2 in October 2008. With the collimator standard resolution, the pixel size was 0$\farcs$125 and 0$\farcs$25 on the sky, respectively. The FORS2 CCD was a 4k$\times$4k MIT mosaic, used with binning 2$\times$2 corresponding to a pixel size of 0$\farcs$25 on the sky.

\section{Data reduction and measurements}
\label{sec:reduc}

\begin{table*}[t]
\caption[ ]{Observed standard stars}
\label{tab:std}
\begin{tabular}{lll}
\hline\hline
Date & Polarized & Unpolarized\\ 
yyyy-mm-dd & &\\
\hline\\
2008-05-08  & Ve6$-$23, BD$-$12$\degr$5133                     & WD0752$-$676, WD2149$+$021  \\
2008-05-09  & Ve6$-$23,                                        & WD0752$-$676 \\
2008-05-10  & Ve6$-$23, BD$-$12$\degr$5133, BD$-$14$\degr$4922 & WD0752$-$676 \\
2008-10-29  & NGC2024$-$1                                      &  \\
2008-10-30  & NGC2024$-$1                                      &  \\
2015-05-15  &                                                  & WD2039$-$202  \\
2015-06-10  &                                                  & WD1344$+$106  \\
2015-06-18  & BD$-$12$\degr$5133                               &  \\
2015-07-10  & BD$-$12$\degr$5133                               & WD2149$+$021 \\
\hline
\end{tabular}
\tablebib{\citet{2007Fossati}} 
\end{table*} 

The $q$ and $u$ Stokes parameters were computed from the ratios of the integrated intensities of the orthogonally polarized images of the object, measured for the four different orientations of the HWP. They were calculated with respect to the instrumental reference frame according to
%
\begin{eqnarray}
q & = & \frac{R_q - 1}{R_q + 1}  \hspace{0.5cm} \mbox{where} \hspace{0.5cm} 
R_q^2  = \frac{I_{\srm 0}^{\rm u}/I_{\srm 0}^{\rm l}}
       {I_{\srm 45}^{\rm u}/I_{\srm 45}^{\rm l}},\;\;\; \rm{and} \nonumber\\
 & &\\
u & = & \frac{R_u - 1}{R_u + 1}  \hspace{0.5cm} \mbox{where} \hspace{0.5cm} 
R_u^2  = \frac{I_{\srm 22.5}^{\rm u}/I_{\srm 22.5}^{\rm l}}
       {I_{\srm 67.5}^{\rm u}/I_{\srm 67.5}^{\rm l}},\nonumber 
\end{eqnarray}
where $I^{\rm u}$ and $I^{\rm l}$ refer to the intensities (electron counts) integrated over the upper and lower images of the object, respectively. This combination of measurements from the four frames secured with different HWP orientations removes most of the instrumental polarization, and corrects for the effects of image distortions that can be generated by the HWP \citep{1989diSerego,1999Lamy}. The intensity measurements were performed using the procedures described in \citet{1999Lamy} and \citet{2005Sluse}. Specifically, the positions of the  upper and lower images were measured at subpixel precision by fitting two-dimensional Gaussian profiles. The intensities were then integrated in circles centered on the upper and lower images, and the Stokes parameters computed for increasing values of the aperture radius. Since the Stokes parameters are most often found to be stable for increasing aperture radii, we adopted a fixed aperture radius of $3.0 \times [(2 \ln 2)^{-1/2}\, \rm{HWHM}],$ where HWHM is the mean half-width at half-maximum of the two-dimensional Gaussian profile. In a few cases, the Stokes parameters strongly fluctuated when the aperture radius was changed, which made these measurements unreliable. The uncertainties $\sigma_q$ and $\sigma_u$ were estimated by computing the errors on the intensities $I^{\rm u}$ and $I^{\rm l}$ from the read-out noise and the photon noise in the object and the sky background, and then by propagating these errors. Typical uncertainties are around 0.1 \% for either $q$ or $u$.

\begin{table*}[t]
\caption[ ]{Polarization of distant stars}
\label{tab:halo}
\begin{tabular}{lccccccccccr}
\hline\hline
Reference-number & RA & DEC &  Distance &  Obs. Date & $q$ & $u$ & $p$  & $\sigma_p$ & $p_0$ & $\theta$ & $\sigma_{\theta}$  \\
     & h m s    &  $\degr$ $\arcmin$ $\arcsec$ & kpc & & \% & \% & \% & \% & \% & \degr & \degr \\
\hline \\
Beers-745           &  11 15 47.10 & $-$17 55 56.7  & 19.6  &  2008-05-08 &  $+$0.01  &  $-$0.04  &  0.04  &  0.05  &  0.00  &  -  &  - \\
Beers-747           &  11 16 46.59 & $-$16 59 52.4  & 13.8  &  2008-05-08 &  $-$0.06  &  $+$0.02  &  0.06  &  0.07  &  0.00  &  -  &  - \\
Beers-748           &  11 16 59.95 & $-$19 33 57.7  & 21.7  &  2008-05-10 &  $-$0.04  &  $-$0.07  &  0.08  &  0.08  &  0.00  &  -  &  - \\
Beers-752           &  11 19 32.13 & $-$17 05 11.5  & 19.4  &  2008-05-10 &  $+$0.02  &  $+$0.34  &  0.34  &  0.09  &  0.33  &  43 &  8 \\
Clewley-CF789-045   &  12 43 23.35 & $-$04 11 29.8  & 15.1  &  2008-05-09 &  $-$0.08  &  $-$0.06  &  0.10  &  0.07  &  0.08  & 107 & 25 \\
Clewley-CF789-041   &  12 44 20.10 & $-$03 21 38.1  & 18.2  &  2008-05-10 &  $-$0.06  &  $+$0.12  &  0.14  &  0.08  &  0.12  &  58 & 19 \\
\hline
\end{tabular}
\tablebib{\citet{2000Beers,2004Clewley}} 
\end{table*} 

A zero-point angle offset was then applied to the normalized Stokes parameters $q$ and $u$ in order to convert the polarization angle measured in the instrumental reference frame into the equatorial reference frame. For the V filter, the offset was 1.8$\degr$, according to the FORS user manual. This angle offset was checked using polarized standard stars (Table~{\ref{tab:std}}). For all standard stars, the measured values of the polarization angles corrected with that offset are within 1\degr\ of their nominal values. The polarization of unpolarized standard stars (Table~{\ref{tab:std}}) is around 0.10 $\pm$ 0.05 \% for all runs, indicating that the residual instrumental polarization is small at the center of the field.

Then, the polarization degree was computed using $p = (q^2+u^2)^{1/2}$ and the associated error $\sigma_p \simeq \sigma_q \simeq \sigma_u$.  The debiased value $p_{0}$ of the polarization degree was obtained using the \citet{1974Wardle} estimator, which is a reasonably good estimator of the true polarization degree \citep{1985Simmons}. The polarization position angle $\theta$ was obtained by solving the equations $q = p\cos 2\theta$ and $u = p \sin 2\theta$. The uncertainty of the polarization position angle $\theta$ was estimated from the standard \citet{1962Serkowski} formula, where the debiased value $p_{0}$ was conservatively used instead of $p$, that is, $\sigma_{\theta} = 28.65\degr \, \sigma_p / p_{0}$ \citep[see also][]{1974Wardle}. 

As in \citet{2017Hutsemekers}, we secured the V-band polarization of a few distant stars ($d > $ 10 kpc) to check the magnitude of the interstellar polarization in the direction of our targets. These measurements are reported in Table~\ref{tab:halo}. All these stars have low polarization. Although the sample is small, this confirms that on average, contamination by interstellar polarization is essentially negligible for quasars at high galactic latitudes ($|b_{\rm gal}| > 30 \degr$) and with polarization degrees higher than 0.6\%  \citep{1990Berriman,2000Lamy,2005Sluse,2017Pelgrims}.

\section{Polarization data}
\label{sec:data}

The full Table~3, available at the Strasbourg astronomical Data Center (CDS), contains 87 polarization measurements obtained for 86 quasars (63 measurements in May 2008, 4 in October 2008, and 20 in 2015). Unreliable measurements were discarded (measurements for which no stable value of the Stokes parameters could be secured; see Sect.~\ref{sec:reduc}).   Thirty-seven quasars have $p \geq 0.6\%$, 9 have $p \geq 2\%$, and 1 has $p \geq 10\%$.  Column~(1) gives the quasar name from the NASA/IPAC Extragalactic Database (NED), Cols.~(2) and~(3) the equatorial coordinates (J2000), Col.~(4) the redshift $z$, Col.~(5) the filter, and Col.~(6) the date of observation (year-month-day). Columns~(7) and~(8) give the normalized Stokes parameters $q$ and $u$ in percent. The normalized Stokes parameters are expressed in the equatorial reference frame.  Columns~(9) and~(10) give the polarization degree $p$ and its error $\sigma_p$ in percent. Column~(11) gives the debiased polarization degree $p_0$ in percent. Columns~(12) and~(13) give the polarization position angle $\theta$ east-of-north and its error $\sigma_{\theta}$, in degree. When $p < \sigma_p$, the polarization angle is undefined and its value set to 999.

The five objects with $p \geq 3\%$ are reported in the excerpt of Table~\ref{tab:qsos}. For all of them, polarization measurements are secured for the first time. SDSS J112738.76+013537.9 is a BAL quasar \citep{2006Trump}. PKS 2054-377, WISE J121043.78-275858.9, PKS 1336-237, and WISE J115217.19-084103.1 are Parkes radio sources. The last two objects also belong to the Fermi Gamma-ray Space Telescope source catalogue \citep{2015Acero}.

\begin{sidewaystable}
\caption[ ]{Polarization of quasars}
\label{tab:qsos}
\centering   
\begin{tabular}{lcccccrrrrrrr}
\hline\hline
Name & RA    & DEC                         & $z$ & Filter & Obs. Date & $q$ & $u$ & $p$  & $\sigma_p$ & $p_0$  & $\theta$ & $\sigma_{\theta}$   \\
     & h m s & $\degr$ $\arcmin$ $\arcsec$ &     &        &           & \%  & \%  & \%   & \%         & \%  & \degr    & \degr             \\
\hline \\
EIS J033252.61-273846.5          & 03 32 52.60 & $-$27 38 46.2 &  1.023300 &  V &  2008-10-30 &     0.78 &  $-$0.14 &   0.79 &   0.85 &   0.00 &   999 &  999    \\
SDSS J112738.76+013537.9         & 11 27 38.76 & $+$01 35 38.0 &  2.014061 &  V &  2008-05-09 &  $-$6.75 &  $-$2.32 &   7.14 &   0.17 &   7.14 &   100 &    1    \\
WISE J115217.19-084103.1         & 11 52 17.21 & $-$08 41 03.3 &  2.370000 &  V &  2008-05-08 &     6.05 & $-$11.78 &  13.24 &   0.10 &  13.24 &   149 &    1    \\
WISE J121043.78-275858.9         & 12 10 43.61 & $-$27 58 54.6 &  0.828000 &  V &  2008-05-09 &  $-$7.80 &     4.36 &   8.94 &   0.12 &   8.94 &    75 &    1    \\
PKS 1336-237                     & 13 39 01.75 & $-$24 01 14.0 &  0.657000 &  V &  2008-05-09 &  $-$9.26 &     0.16 &   9.26 &   0.10 &   9.26 &    90 &    1    \\
PKS 2054-377                     & 20 57 41.60 & $-$37 34 03.0 &  1.071000 &  V &  2008-05-09 &     2.42 &  $-$3.48 &   4.24 &   0.33 &   4.23 &   152 &    2    \\ 
\hline
\end{tabular}
\tablefoot{This table gives the polarization measurements for six quasars. It contains the five objects with $p \geq 3\%$. The complete table is available electronically at the CDS. }
\end{sidewaystable}

\begin{acknowledgements}
This research has made use of the NASA/IPAC Extragalactic Database (NED), which is operated by the Jet Propulsion Laboratory, California Institute of Technology, under contract with the National Aeronautics and Space Administration. 
\end{acknowledgements}

\bibliographystyle{aa}
\bibliography{references}

\clearpage
\includepdf[pages=1]{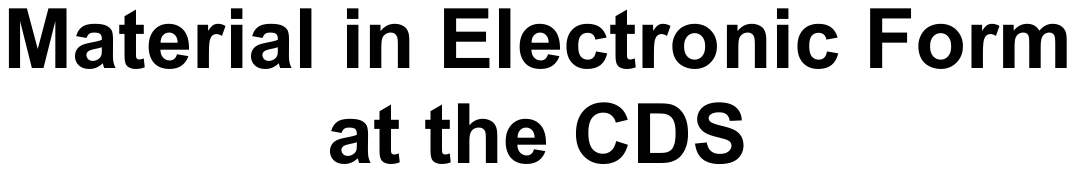}
Table 3 : Polarization of 86 quasars
\includepdf[pages=1]{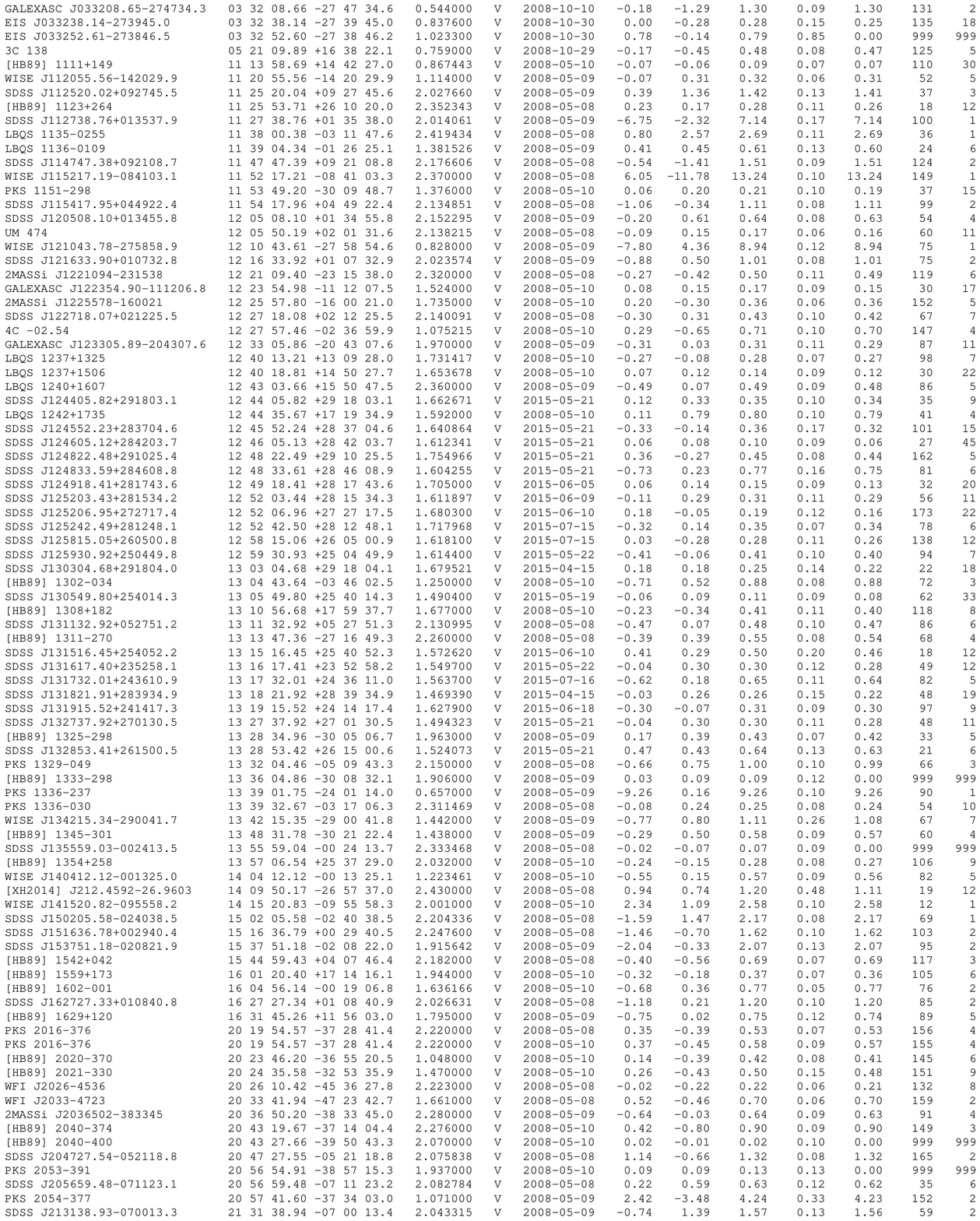}

\end{document}